\documentclass[iop]{emulateapj}
\usepackage{natbib}
\newcommand\etal{et al.}
\newcommand\psra{PSR J1628$-$3205}
\newcommand\psrb{PSR J1301+0833}

\slugcomment{To Appear in The Astrophysical Journal}

\shorttitle{Optical Counterparts of Millisecond Pulsars}
\shortauthors{Li, Halpern, Thorstensen}

\begin{document}

\title{Optical Counterparts of Two {\it Fermi} Millisecond Pulsars:\\
\psrb\ and \psra}

\author{Miao Li, Jules P. Halpern}
\affil{Department of Astronomy, Columbia University, 550 West 120th Street, New York NY, 10027, USA; miao@astro.columbia.edu}
\author{John R. Thorstensen}
\affil{Department of Physics and Astronomy, Dartmouth College, Hanover NH, 03755, USA}

\begin{abstract}
Using the 1.3m and 2.4m telescopes of the MDM Observatory, we 
identified the close companions of two eclipsing millisecond
radio pulsars that were discovered by the Green Bank Telescope in
searches of {\it Fermi Gamma-ray Space Telescope} sources,
and measured their light curves.
\psrb\ is a black widow pulsar in a 6.5~hr orbit whose companion star
is strongly heated on the side facing the pulsar.  It varies from 
$R=21.8$ to $R>24$ around the orbit.  \psra\ is a ``redback,''
a nearly Roche-lobe filling system in a 5.0~hr orbit whose
optical modulation in the range $19.0<R<19.4$ is dominated by strong
ellipsoidal variations, indicating a large orbital inclination angle.
\psra\ also shows evidence for a long-term variation of about 0.2 mag,
and an asymmetric temperature distribution possibly due to either
off-center heating by the pulsar wind, or large starspots.
Modelling of its light curve restricts the inclination angle to
$i>55^{\circ}$, the mass of the companion to $0.16<M_c<0.30\,M_{\odot}$,
and the effective temperature to $3560<T_{\rm eff}<4670$~K.  As is the
case for several redbacks, the companion of \psra\ is less dense
and hotter than a main-sequence star of the same mass.

\end{abstract}

\keywords{gamma rays: stars ---pulsars: individual (\psrb, \psra)}

\section{Introduction}

Millisecond pulsars (MSPs) are neutron stars (NSs) that are spinning
faster than any young pulsar.   Shortly after their discovery, the
``recycling'' model was proposed by \cite{alpar82} and \cite{radhakrishnan82}
to explain how MSPs acquire such high rotation velocity: they were previously
spun up by accretion through Roche lobe overflow of their companions in
low-mass X-ray binaries (LMXBs) as the orbit shrinks due to magnetic braking
and/or gravitational radiation.
This scenario is consistent with the preponderance of MSPs in binary systems,
and it also explains the fact that they have much weaker magnetic fields
than ordinary pulsars, since accretion can dissipate and bury magnetic
field \citep{rom90,gep94}.

An interesting subclass of MSPs are the eclipsing radio MSPs, the so-called
``black widows'' and ``redbacks.''  Approximately 53 such systems are
known\footnote{http://apatruno.wordpress.com/about/millisecond-pulsar-catalogue/}
from large-scale radio surveys, deep searches of globular clusters,
and most recently,
radio follow-ups of unidentified {\it Fermi} $\gamma $-ray sources
\citep[e.g.,][]{hessels11, ransom11, ray12}. Black widows and redbacks are
compact binaries with short orbital periods ($\lesssim 1$~day) and small
companion masses, $\approx 0.01-0.05 M_{\sun}$ for black widows, and
$\approx 0.1-1 M_{\sun}$ for redbacks.
Radio eclipses occur when the pulsar is at superior conjunction. The eclipsing
plasma is usually more extended than the Roche lobe of the
companion, which motivated
the theory that the companion is being ablated by the high-energy particles
and/or photons from the pulsar \citep{phinney88}. Indeed, optical and X-ray
fluxes show orbital modulation indicating that the side of the companion
facing the pulsar is being heated
\citep[e.g.][]{rom11, kong12, breton13, gentile14, bog11, bog14a, bog14b}.  

\begin{deluxetable*}{ccccc}
\tabletypesize{\small}
\tablewidth{0pt}
\tablecaption{Log of Observations}
\tablehead{
\colhead{Object} & \colhead{Telescope} & \colhead{Date (UT)} &
\colhead{Time (UT)} & Orbital Phase ($\phi$)\tablenotemark{a}
}
\startdata
\psrb  &  1.3m  &  2013 April 4  &  04:29--08:50  &  0.76--0.13  \\
``     &  2.4m  &  2014 May 26   &  03:45--05:40  &  0.73--0.02  \\
``     &  2.4m  &  2014 May 27   &  03:32--05:54  &  0.43--0.72  \\
``     &  2.4m  &  2014 May 28   &  04:48--05:20  &  0.23--0.31  \\
\hline
\psra  &  1.3m  &  2013 April 7  &  09:30--12:18  &  0.57--0.13  \\
``     &  1.3m  &  2013 July 7   &  05:35--07:02  &  0.99--0.25  \\
``     &  2.4m  &  2014 May 26   &  05:58--09:36  &  0.88--0.60  \\
``     &  2.4m  &  2014 May 27   &  06:04--09:37  &  0.70--0.40  \\
``     &  2.4m  &  2014 May 28   &  05:34--09:28  &  0.41--0.18
\enddata
\tablenotetext{a}{Phase zero corresponds to the ascending node of the pulsar.}
\label{tab:log}
\end{deluxetable*}

\begin{figure*}
\begin{center}
\includegraphics[width=0.47\textwidth]{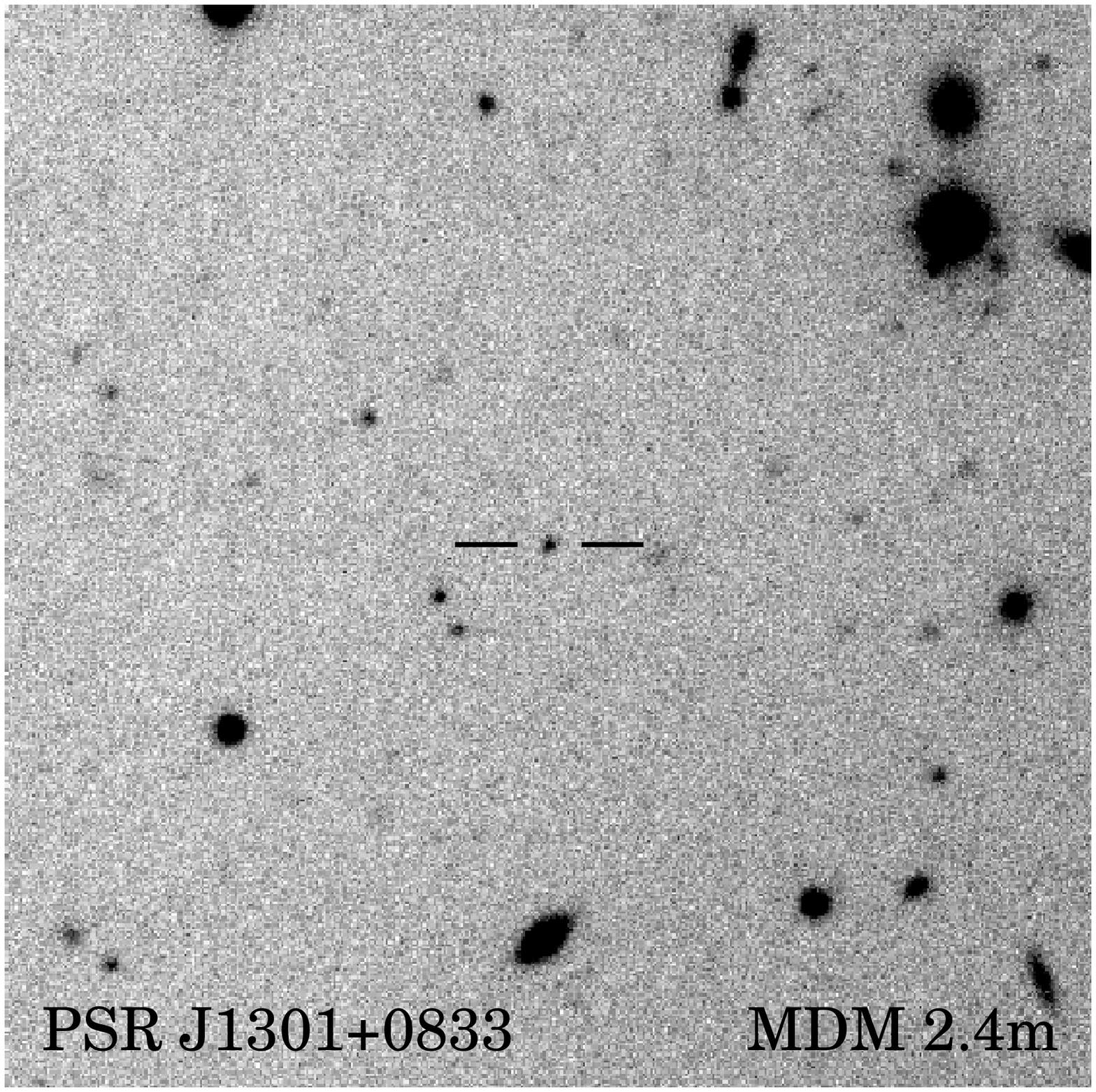}
\includegraphics[width=0.47\textwidth]{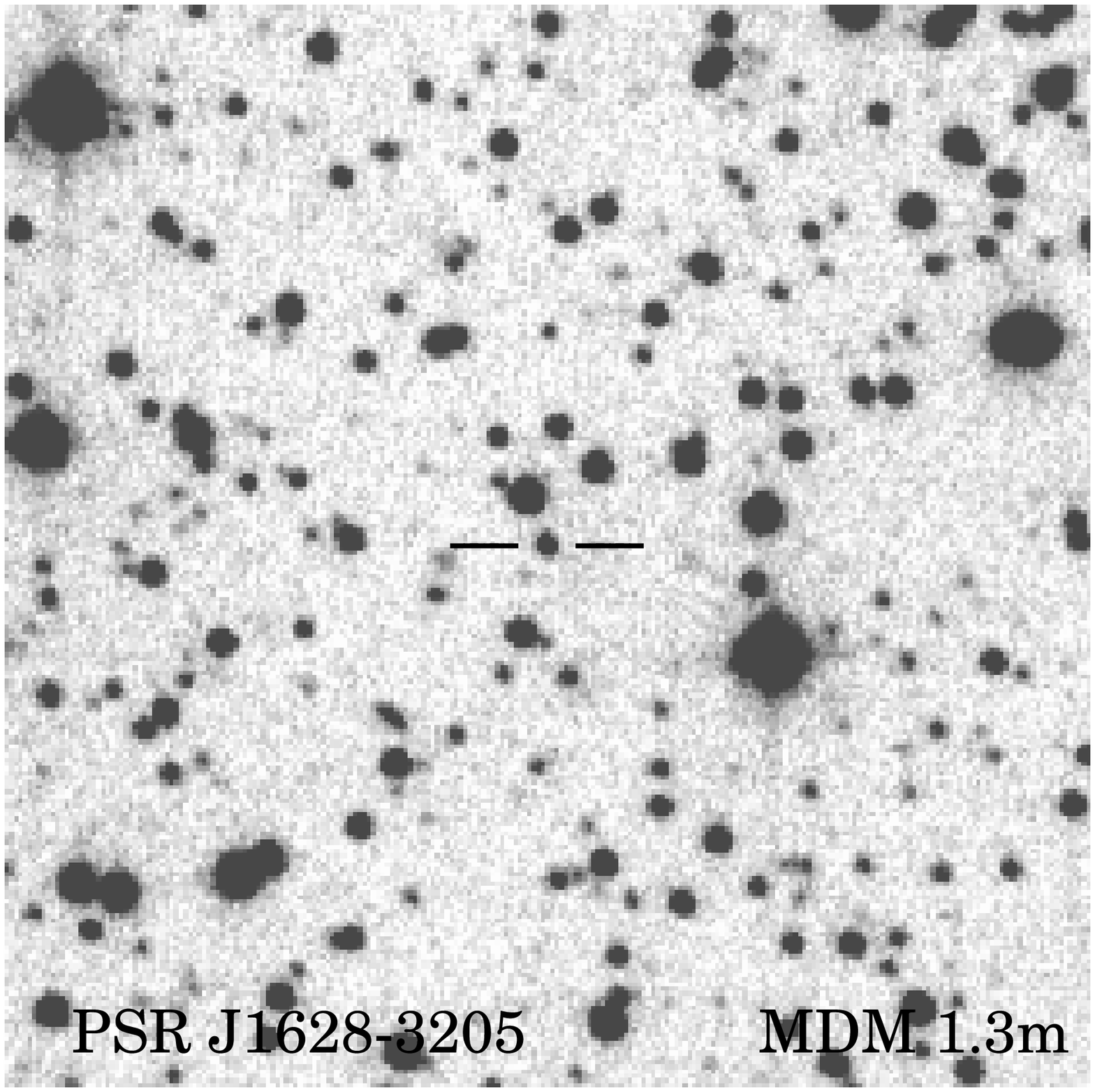}
\end{center}
\caption{Finding charts for \psrb\ (left) and \psra\ (right),
taken from MDM $R$-band images.
Positions are (J2000.0) R.A.=$13^{\rm h}01^{\rm m}38.\!^{\rm s}26$,
decl.=$+08^{\circ}33^{\prime}57.\!^{\prime\prime}5$
and R.A.=$16^{\rm h}28^{\rm m}07.\!^{\rm s}02$,
decl.=$-32^{\circ}05^{\prime}48.\!^{\prime\prime}7$, respectively.
The size is $2^{\prime}\times2^{\prime}$.  North is up, and east is to the left.
\vspace{0.1in}
}
\label{fig:charts}
\end{figure*}

It is unclear how black widow and redback binaries arrived at their current
state, given the complexity of mass transfer and ablation. We do not even
know if redbacks evolve into black widows, or if they have completely
different histories.  Neither fall on the normal binary evolution sequences;
on the diagrams of orbital period versus companion mass and luminosity
versus temperature \citep[e.g., Figure 2 of][]{pod02},
a fair fraction of redbacks
appear in the gap between the evolutionary tracks of long- and short-period
systems, whereas black widows occupy a completely different region.
Mass loss due
to irradiation may have played a role \citep{chen13,ben14}.  Ablation has also
been suggested to explain the existence of isolated MSPs outside of
globular clusters \citep[e.g.][]{kluzniak88}, but it may be difficult
to heat the photosphere enough to achieve the required mass flux
to completely evaporate the companion \citep{eichler88}.
More multi-wavelength observations and physical understanding of pulsar
wind interaction with stellar material are needed to resolve these issues. 

Recently, transitions from accretion mode to radio MSPs have been found
in several redback
systems (PSR J1023+0038, PSR J1824$-$2452I, and XSS J12270$-$4859) that
decline in X-ray luminosity and exhibit millisecond radio pulsations
\citep{archibald09, papitto13, roy14}. 
Optical spectra also show that the blue continuum and double-peaked emission
lines from accretion disks have been replaced by pure stellar absorption
\citep[e.g.][]{thorstensen05, wang09, bas14}.  The opposite transitions have
also been observed in PSR J1824$-$2452I \citep{papitto13} and PSR J1023+0038
\citep{stappers13, halpern13}. 
X-ray luminosities in the accretion state vary by orders of magnitude,
from $>10^{36}$~erg~s$^{-1}$, which is similar to a typical LMXB, to a mild
increase from the MSP mode \citep{linares14}.  The fast transitions and
variations of luminosity reveal that the binary interaction is complex,
and the mass transfer rate is not always high enough for the accretion
disk to reach the NS and spin it up. With multi-wavelength monitoring,
we expect to see more transitions between accretion- and rotation-powered
states. 

Black widows and redbacks are ideal systems to measure NS mass,
since the orbital radial
velocity of pulsar and companion can be measured separately through radio
and optical Doppler shifts, respectively, and light-curve modelling can 
constrain the inclination angle of the orbit. Of special interest is that
these NSs tend to be more massive due to the previous long accretion
phases. \cite{romani12}, for example, have found the black widow
PSR J1311$-$3430
very likely to have $M_{\rm NS} > 2.1 M_\sun$. This probe of the upper limit
of NS mass helps to constrain the equation of state for super-dense matter,
enabling us to explore fundamental physics under extreme conditions that
are unachievable on the earth \citep[][and references therein]{lattimer07}.

Here we report optical identifications and light curves 
of two binary MSPs that were discovered in a radio survey of {\it Fermi}
sources using the Green Bank Telescope (PI: S. Ransom) and subsequently
timed with the GBT and the Westerbork Synthesis Radio Telescope
to determine their orbital and spin-down parameters
(J. Hessels et al., in preparation).
\psrb\ is a 1.84~ms black widow in a 6.5~hr orbit
\citep{ray12}, while \psra\ is a 3.21~ms redback in a 5.0~hr orbit.
We use the light
curve of \psra, together with the pulsar orbital ephemeris, to constrain
some properties of the system. The paper is organized as follows: we
describe our observations and data reduction in Section~\ref{f:obs}.
In Section~\ref{f:model} we describe the light curve fitting model
and present its results for \psra.  We discuss our findings
in Section~\ref{f:discuss}, and summarize our conclusions
in Section~\ref{f:conclude}.

\begin{figure}[t]
\begin{center}
\includegraphics[width=0.47\textwidth]{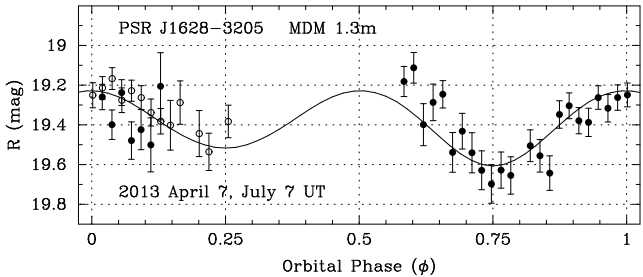}
\end{center}
\vspace{-0.1in}
\caption{$R$-band light curve as a function of orbital phase for \psra.
Filled circles are from 2013 April 7, and open circles from 2013 July 7,
both from the MDM 1.3m telescope.  Phase zero corresponds to the
ascending node of the pulsar, so $\phi=0.75$ is the superior
conjunction of the companion star. The curve is a model of ellipsoidal
modulation described in Section~\ref{f:model}, with $d=1.24$~kpc,
$M_{NS}=1.4\,M_{\odot}$, $M_c=0.167\,M_{\odot}$, $i=75^{\circ}$, and 
$T_{\rm eff}=4130$~K.
}
\vspace{0.1in}
\label{fig:psrj1628_1.3m}
\end{figure}

\section{Observations}
\label{f:obs}

All of our data are from $R$-band images obtained at the MDM Observatory's
1.3m McGraw-Hill telescope or 2.m Hiltner telescope on Kitt Peak
during three observing runs in 2013 and 2014.  A log of the time-series
observations used in this paper is presented in Table~\ref{tab:log}. 
The detector was the thinned, backside illuminated SITe CCD ``Templeton''.
It has $1024\times1024$ pixels, with a scale of
$0.\!^{\prime\prime}509$ pixel$^{-1}$ on the 1.3m,
and $0.\!^{\prime\prime}275$ pixel$^{-1}$ on the 2.4m.  All exposures were
300~s, and readout time was $\approx 30$~s.  The runs were timed
to eventually cover all orbital phases of each pulsar.
The data were reduced using standard procedures in IRAF, including
bias-subtraction, flat-fielding with twilight exposures, and
differential aperture photometry with respect to isolated
comparison stars using {\it phot}.  Photometric calibration of the
comparison stars was done using \citet{landolt92} standard stars
observed immediately before or after the target pulsars.
Heliocentric correction was applied to the light curves.

\subsection{\psra}
\label{f:obs1}

\psra\ is the only one of a group of six
redback systems in the Galactic field from \citet{roberts13} that
that remains to be identified optically.
Optical studies have revealed that redback companion stars
are hot, with $T_{\rm eff}\approx5000-6000$~K, and fill a large fraction
of their Roche lobes \citep{breton13, crawford13, schroeder14}.  An exception
is PSR J1816+4510, which is significantly hotter, $T_{\rm eff}\approx16,000$~K,
and underfills its Roche lobe \citep{kap12,kap13}.  Existing data
suggest they are metal-rich and non-degenerate, consistent with the expectation
that the outer layers have been ablated and the remnant stars are bloated by
heating, but their exact nature is still not clear. 
The minimum mass of the companion to \psra\ is $M_c=0.16\,M_{\odot}$ if
$M_{NS}=1.4\,M_{\odot}$.

The pulsar dispersion measure (DM) of 42.1~pc~cm$^{-3}$ for \psra\
\citep{ray12} corresponds to a distance $d=1.24$~kpc according to
the \citet{cor02} model of the Galactic distribution of free electrons.
We will assume a nominal uncertainty of $\pm30\%$ on the 
distance for the purpose of modelling in Section~\ref{f:model},
but we will also allow for an error of
a factor of $\sim2$, which sometimes occurs, especially
at high Galactic latitude \citep[e.g.,][]{rob11,del12}.
The 0.1--100~GeV luminosity of the {\it Fermi} source 2FGL J1628.3$-$3206
at this distance is $\approx2.1\times10^{33}$ erg~s$^{-1}$ (assumed isotropic),
which is 16\% of its apparent spin-down luminosity,
$\dot E = 4\pi^2 I \dot P/P^3 = 
1.29\times10^{34}$ erg~s$^{-1}$ from the timing parameters
(S. Ransom, private communication).  This is a typical $\gamma$-ray
efficiency for MSPs. Here we have assumed the moment of
inertia $I=1\times10^{45}$ g~cm$^2$, which could be uncertain by
a factor of 2.
The precise radio timing position coincides
with a faint star on the digitized sky survey, with magnitudes
$B2=21.06, R2=18.61$ in the USNO--B1.0 catalog \citep{mon03}.
In images obtained on the 1.3m (Figure~\ref{fig:charts})
we refined its optical position to (J2000.0)
R.A.=$16^{\rm h}28^{\rm m}07.\!^{\rm s}02$,
decl.=$-32^{\circ}05^{\prime}48.\!^{\prime\prime}7$
in the USNO--B1.0 reference frame.

\begin{figure*}
\begin{center}
\includegraphics[width=0.86\textwidth]{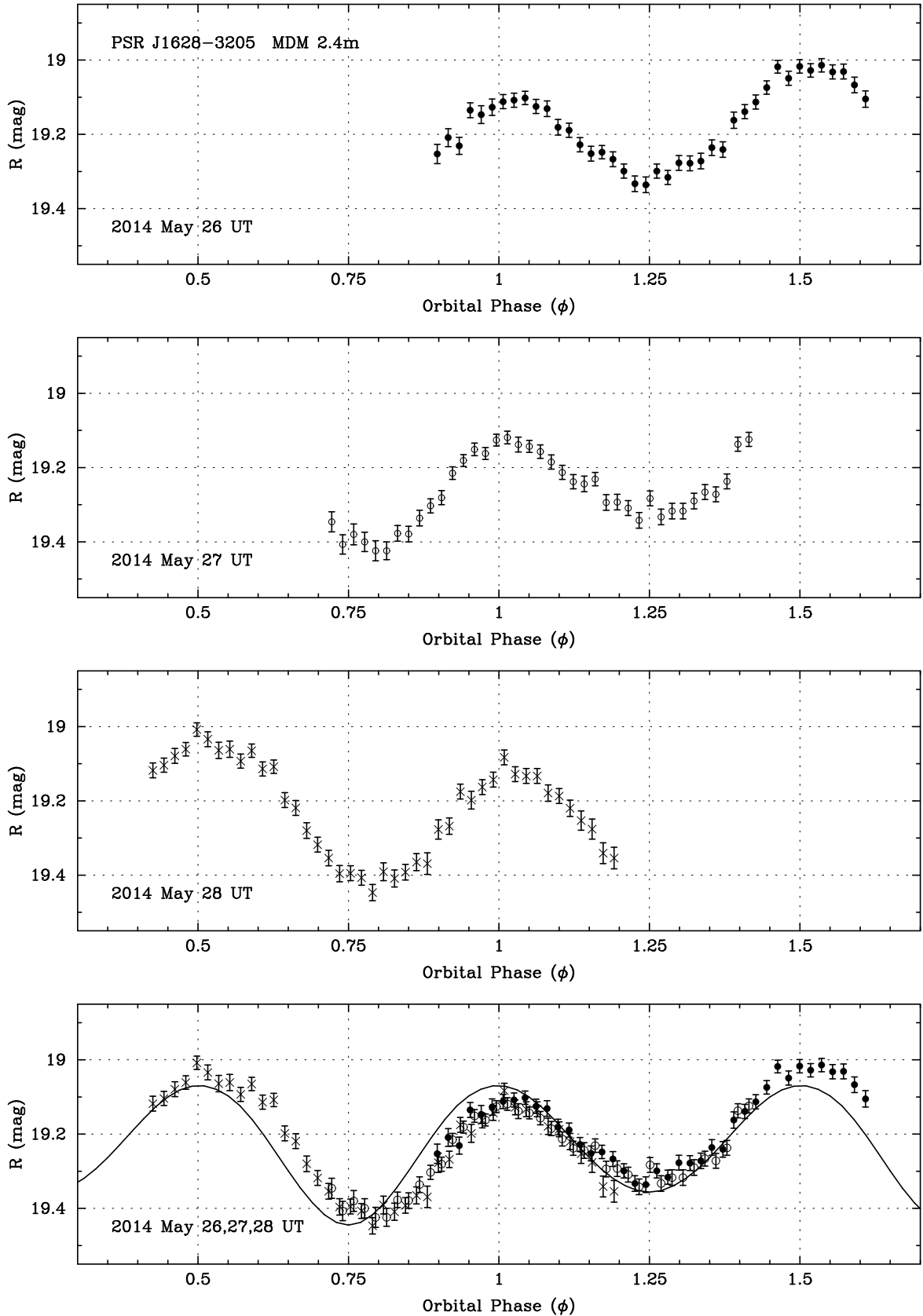}
\end{center}
\caption{$R$-band light curves as a function of orbital phase for \psra,
obtained on three consecutive nights in 2014 May, on the MDM 2.4m.
The bottom panel shows the points from the three nights superposed,
retaining their different symbols from the individual nights.
The curve is the representative ellipsoidal model described in
Section~\ref{f:model},  with $d=1.24$~kpc,
$M_{NS}=1.4\,M_{\odot}$, $M_c=0.167\,M_{\odot}$, $i=75^{\circ}$, and 
$T_{\rm eff}=4230$~K.
}
\label{fig:psrj1628_2.4m}
\end{figure*}

Due to the southerly declination of \psra, as well as partly cloudy conditions,
we were only able to observe $\approx 68\%$ of the phase of its 6.0~hr
orbit using the 1.3m in 2013.  Nevertheless, the coverage was complete
enough to reveal that there are two flux minima and two maxima per orbit
(Figure~\ref{fig:psrj1628_1.3m}),
a modulation consistent with ellipsoidal deformation of a nearly
Roche-lobe filling star in a highly
inclined orbit.  The basic manifestations of ellipsoidal
variations are 1) equal maxima at quadrature phases (0.0 and 0.5)
when the projected area of the tidally distorted star is largest,
and 2) unequal minima at phases 0.25 and 0.75
due to the larger effects of gravity darkening
and limb darkening when viewing the L1 point of the companion
star at $\phi=0.75$.   Fitting of the light curve including these
effects will be presented in Section~\ref{f:model}.
Evidently, heating of the companion by the
pulsar wind, which should contribute most at $\phi=0.75$,
is not a major effect in this system.

In 2014 May, we obtained additional light curves of \psra\ using the
MDM 2.4m telescope (Figure~\ref{fig:psrj1628_2.4m}).  As before,
all observations were obtained at high airmass, sec~$z=2.3-2.9$.
Although it took three nights to cover all
orbital phases, the substantial overlap in phase among the nights shows
that the light curve is largely stable over this time interval.  The
higher precision of the 2.4m data establishes three interesting
properties of the light curve.  First, the two maxima,
at $\phi=0.0$ and 0.5, are not quite equal, differing by $\approx 0.1$~mag.
This implies that the temperature distribution on the companion is
not symmetric about the axis connecting the stars.  The trailing
side of the companion is brighter. Second, the deeper minimum in the
light curve, which is expected to fall exactly at $\phi=0.75$, lags by
about 0.05 in phase, while the minimum at $\phi=0.25$ is not displaced.
These two effects could indicate heating from the pulsar that is not symmetric
about the line between the stars, but stronger on the trailing side of the
companion.  Alternatively, there could be an intrinsic temperature
distribution resulting from large starspots.  The third interesting
result is an increase in brightness from
the previous year (Figure~\ref{fig:psrj1628_1.3m}) by an amount
that ranges from $0.1-0.2$~mag around the orbit.  This could be due to an
increase in the radius or temperature of the photosphere,
perhaps from enhanced heating
by the pulsar.  Historical detections of \psra\ on the Digitized Sky
Survey plates are broadly consistent with our CCD
magnitudes, but they are too uncertain to be evaluated for additional
evidence of variability.

\begin{figure}[t]
\begin{center}
\includegraphics[width=0.33\textwidth,angle=270.]{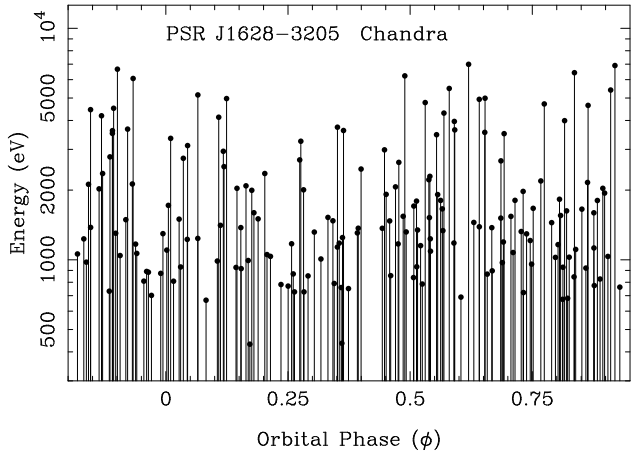}
\vspace{-0.1in}
\end{center}
\caption{Energies and arrival times in orbital phase of 176 photons
detected from \psra\ by {\it Chandra} on 2012 May 2 (ObsID 13725).
Each point is one photon.
}
\label{fig:chandra}
\end{figure}

We note that the magnitude scale in
Figures~\ref{fig:psrj1628_1.3m} and \ref{fig:psrj1628_2.4m} may
have a systematic error of $\sim 0.05$~mag due to the difficulty
of photometric calibration at large airmass.  Nevertheless, the three
effects described in the previous paragraph, which rely only on
differential photometry, have been checked with several comparison stars,
and they appear to be reliable.

We were able to test for proper motion of \psra\ by comparing
its position on our CCD images with a UK Schmidt IV-N plate that
was taken on 1980 July 16.  The latter provides the best detection
of \psra\ of any of the digitized sky survey plates.  Employing
a surrounding grid of 29 stars with positions and proper motions
in the USNO--B1.0 catalog as astrometric references, we calculate
proper motion in Right Ascension and declination as
$\mu_{\alpha}\,{\rm cos}\,\delta=+8\pm9$ mas~yr$^{-1}$ and
$\mu_{\delta}=-16\pm9$ mas~yr$^{-1}$, respectively.  The errors
are dominated by the accuracy of centroiding the object
on the digitized plate image, which we take to be $0.\!^{\prime\prime}3$
in each coordinate.
Although only a marginal detection of motion, this corresponds to a
tangential velocity of $\sim100$ km s$^{-1}$ at $d=1.2$~kpc,
which is relevant to the estimation of the true spin-down
rate of the pulsar as discussed in Section~\ref{f:discuss}.

An observation of \psra\ with the {\it Chandra X-ray Observatory}
ACIS-S3 CCD detected $\approx 176$ photons, ranging in energy
from 0.4--7 keV, from a point source at 
(J2000.0) R.A.=$16^{\rm h}28^{\rm m}07.\!^{\rm s}00$,
decl.=$-32^{\circ}05^{\prime}48.\!^{\prime\prime}9$,
consistent with our optical position and the radio timing position.
The 20~ks exposure on 2012 May 2 (ObsID 13725) spanned 1.25 orbits
of the pulsar.  In Figure~\ref{fig:chandra} we show the arrival times
and energies of those photons as a function of orbital phase.  There is
no apparent orbital modulation of count rate or photon energy, indicating
that the X-rays are probably coming from the pulsar, and not an intra-binary
shock near the companion star (where they could be partly occulted by the
companion). If so, the hard X-rays (above 2~keV) indicate a
preponderance of non-thermal magnetospheric emission over
thermal emission from heated polar caps, which is unusual for MSPs
of the spin-down power of \psra.

\subsection{\psrb}
\label{f:obs2}

\begin{figure}
\begin{center}
\includegraphics[width=0.47\textwidth]{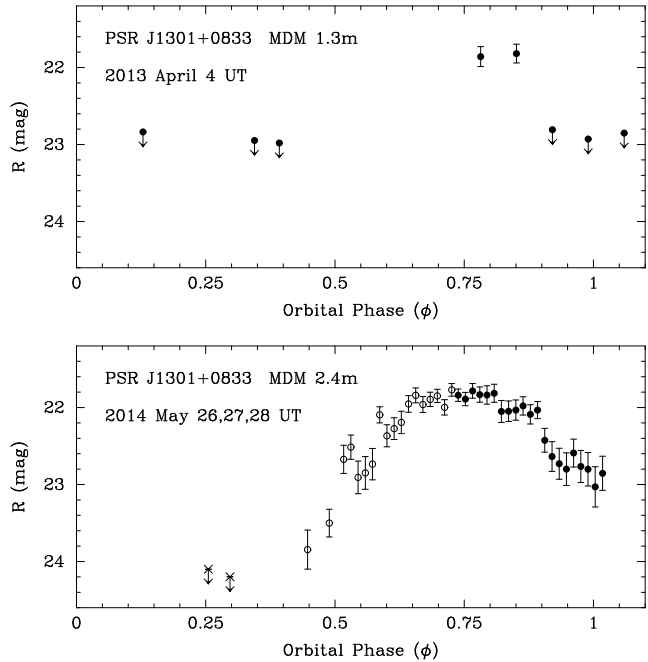}
\end{center}
\vspace{-0.1in}
\caption{$R$-band light curve as a function of orbital phase for \psrb,
showing a heating effect peaking at phase 0.75, the superior conjunction
of the companion star.
Top: Detections and upper limits from the MDM 1.3m telescope are from
the sums of between three and five consecutive 300~s exposures.
Bottom: Data obtained on three consecutive nights in 2014 May on the MDM 2.4m.
The symbols correspond to the same dates as in Figure~\ref{fig:psrj1628_2.4m}.
The four lowest points are from the sums of three consecutive 300~s exposures;
the remainder are individual exposures.
}
\vspace{0.1in}
\label{fig:psrj1301}
\end{figure}

The minimum mass of the companion to \psrb\ is $M_c=0.024\,M_{\odot}$
if $M_{NS}=1.4\,M_{\odot}$, making this a likely black-widow system.
The DM of 13.2~pc~cm$^{-3}$ for \psrb\ \citep{ray12},
corresponds to $d\approx0.67$~kpc.
The 0.1--100~GeV luminosity o f the {\it Fermi} source 2FGL J1301.5+0835
at this distance is $\approx4.5\times10^{32}$ erg~s$^{-1}$ (assumed isotropic),
which is 0.9\% of its apparent spin-down luminosity,
$\dot E\approx5\times10^{34}$ erg~s$^{-1}$ from the timing parameters
(F. Camilo, private communication), corrected for proper motion.
In images obtained at the 2.4m
we found a variable star at the precise radio timing position.
It was also detected by the 1.3m, although too faint there to
be studied in detail.  Figure~\ref{fig:charts} shows a finding
chart for \psrb\ from the 2.4m.  Its position in the USNO-B1.0 reference
frame is (J2000.0) R.A.=$13^{\rm h}01^{\rm m}38.\!^{\rm s}26$,
decl.=$+08^{\circ}33^{\prime}57.\!^{\prime\prime}5$.

The $R$-band light curve of \psrb\ is shown in Figure~\ref{fig:psrj1301}.
The most useful data come from three consecutive nights in 2014 May, the
same as were used to study \psra.  Sharing the time between these
targets, we were only able to cover the complete 6.5~hr orbit of \psrb\
by scheduling it in parts over three nights.   The maximum brightness 
of $R=21.8$ occurs at phase 0.75, which is expected for maximal heating
of the companion by the pulsar wind.  There is a hint of flaring behavior
on the rising side of the light curve, from
$0.50<\phi<0.58$, but additional data with higher time resolution
would be needed to confirm this.  This star is also detected as
SDSS J130138.25+083357.6 with $r=21.86\pm0.12,\ i=21.46\pm0.13$.
Evidently the Sloan Digital Sky Survey
caught the star near maximum light.

The night of May 28 was partly
cloudy during the observations of this target, resulting in no coverage of
phases $0.05-0.2$ and $0.35-0.4$.  However, these phases are close to
inferior conjunction of the companion, where it is already obvious from
the upper limits that we did obtain that the intrinsic magnitude of the
star is fainter than $R=24$.
These upper limits were obtained from the sums of three consecutive 300~s
exposures, as were the two faintest detections at $\phi=0.43-0.5$;
the remainder of the points in Figure~\ref{fig:psrj1301} are from
individual exposures.

\section{Light-Curve Modelling}
\label{f:model}

Although the \psra\ system is evidently not accreting, it is apparent
from its large ellipsoidal variations that the secondary star nearly fills
its Roche lobe.  Therefore, we modelled the light curve using the code of
\citet{thorstensen05}, which was designed for the redback PSR~J1023+0038.
It accounts for effects of gravity darkening and limb darkening, in
addition to treating the pulsar as a point-source of radiation that becomes
thermalized in the photosphere of the companion and is re-radiated locally
at the higher effective temperature
needed to carry the added luminosity.  The surface brightness at a
given effective temperature is estimated from stellar atmosphere models.

We modified the code to make predictions for the
Cousins $R$-band, and removed the parts of the code that predict and
fit the color modulation, since we have only one passband.  The program
predicts the brightness of a Roche-lobe filling star using an
approximation to the surface-brightness versus $T_{\rm eff}$ relation. 
For the present use, in addition to adopting values appropriate to the
$R$-band, we extended the temperature range down to
2400~K using M-dwarf data tabulated by \citet{cas08}.  The limb- and
gravity-darkening coefficients were not changed from their standard
values in \citet{thorstensen05}.

First, we estimate the correction for interstellar extinction.
At the DM distance $d=1.24$~kpc and Galactic latitude
of $11.\!^{\circ}5$, \psra\ has a $z$-height of 250~pc, possibly placing
it above most of the Galactic dust layer.  The corresponding
free electron column of $N_{\rm e}=1.3\times10^{20}$~cm$^{-2}$, assuming
a typical ionized fraction of 0.1 \citep{he13},
is accompanied by a neutral column
of $N_{\rm H}=1.3\times10^{21}$~cm$^{-2}$, with an associated
visual extinction of $A_V\approx0.72$ mag according to
the conversion $N_{\rm H}/A_V=1.8\times10^{21}$ cm$^{-2}$~mag$^{-1}$
of \citet{pre95}.  Alternatively, 
the total line-of-sight dust extinction in this direction,
$A_V\approx1.07$ \citep{sch11}, sets an upper limit to the extinction
of \psra.
Scaling to the $R$-band via $A_R/A_V=0.79$, the corresponding estimate
and upper limit are $A_R\approx0.57$ and $A_R\approx0.85$, respectively.
We adopt here $A_R=0.75$ mag for our calculations.

The higher-quality data from the 2.4m demonstrates that,
while the light curve is dominated by ellipsoidal modulation,
ellipsoidal variations alone cannot fit accurately.  This is
because of the unequal maxima, and the fact that the lower minimum
occurs later than the expected phase of 0.75
(see Figure~\ref{fig:psrj1628_2.4m} and Section~\ref{f:obs1}).
Large magnetic spots on the secondary
could plausibly explain the distortions; we did not attempt
to model this.  In any case, the mismatch between the data and
model light curves is mostly systematic.
Because of this, a formal best-fit analysis would necessarily be
misleading, so we used the models only to establish broad 
limits on the system parameters, as follows.

{\it Heating effects.}  If the inward
face of the secondary were heated significantly by the pulsar,
the minimum near phase 0.75 would be boosted, but we see no 
indication of that.  For modeling purposes, we leave the heating 
luminosity from the pulsar set to zero.

{\it Mass range.} We assume that $M_{NS}$ is between
1.4 and $2.6\,M_{\odot}$; the range extends to the high side
because pulsars in these systems are thought to have accreted
significant mass during their spin-up.  Pulsar timing gives
an extremely accurate $a_{NS}\,{\rm sin}\,i$, corresponding to 
$v_{NS}\,{\rm sin}\,i=42.972$ km~s$^{-1}$, so that
for any assumed $M_{NS}$ and inclination $i$, the companion-star
mass $M_c$ is fixed.  The minimum possible companion mass
for our assumed range of NS mass is $M_c=0.161\,M_{\odot}$,
corresponding to $M_{NS}=1.4\,M_{\odot}$ and $i=90^{\circ}$.

{\it Inclination.}  The amplitude of the modulation depends 
primarily on $i$; it also increases slightly at lower assumed $T_{\rm eff}$.  
After examining sample light curves, we concluded that the amplitudes
predicted by models with inclination less than $55^{\circ}$ were simply
too low to match the data, so we adopt $i>55^{\circ}$ as a firm 
limit.  We cannot rule out inclinations as high as $90^{\circ}$.
Note that adding a constant source
of light, or assuming that the secondary underfills its Roche lobe,
would tend to make the modulation smaller, and hence increase
the lower limit on $i$.

{\it Effective Temperature.} This is the ``base'' temperature,
the effective surface temperature in the absence of gravity darkening.
Without colors or spectra, we cannot find the secondary star's 
$T_{\rm eff}$ directly, but we can estimate
it as follows.  The large ellipsoidal modulation implies that the
secondary must come close to filling its Roche lobe.  Assuming
a NS mass and orbital inclination, we can then compute the 
secondary's mass and radius.  Assuming a value of
$T_{\rm eff}$ then fixes the surface brightness, and 
hence the secondary's absolute magnitude; matching to the 
observed (extinction-corrected) flux then gives a distance.  
We can constrain $T_{\rm eff}$ by requiring the derived distance
to lie within the range estimated from the dispersion measure,
nominally within $\pm30\%$ of $d=1.24$~kpc.

For a fixed flux, the highest effective temperature corresponds 
to the smallest stellar angular diameter, i.e., the smallest
stellar radius at the greatest distance.  The smallest
secondary radius corresponds to $M_{NS}=1.4\,M_{\odot}$, $i=90^{\circ}$,
$M_c=0.161\,M_{\odot}$, and the largest distance, $\sim1.6$~kpc; for these
parameters, the observed flux is matched with $T_{\rm eff}=4670$~K
(assuming $A_R=0.75$~mag).  The largest angular diameter --
with $M_{NS}=2.6\,M_{\odot}$, $i=55^{\circ}$, 
$M_c=0.297\,M_{\odot}$, and $d=870$~pc  -- yields 
$T_{\rm eff}=3560$~K.  This range of temperatures corresponds 
to main-sequence spectral types between K2 and M2 \citep{pickles98}; 
the secondary's spectrum most likely resembles a late K star.

However, at $d=1.6$~kpc, the absolute $R$ magnitude of \psra\ is +7.4,
which, if it were an isolated main-sequence star, would correspond to type
K7 with $M=0.6\,M_{\odot}$ and $T_{\rm eff}=4000$~K.  At $d=870$~pc, the absolute
$R$ magnitude is +8.8, which corresponds to type M2 with
$M=0.4\,M_{\odot}$ and $T_{\rm eff}=3550$~K.   Main-sequence stars are more
massive, and generally cooler than we have derived from modelling
the light curve.  The companion to \psra, having $M_c<0.3\,M_{\odot}$,
is less dense than a main-sequence star of its mass, and probably hotter.

If we allow a larger uncertainty in distance, say $620<d<2500$~pc, the
effective temperature limits become 3270~K and 5340~K,
which corresponds to main-sequence spectral types between M3 and G8.
However, the conclusion remains that the companion star is less dense
and hotter than a main-sequence star of the same mass.

A representative model corresponding to the nominal distance of
1.24~kpc is shown in Figure~\ref{fig:psrj1628_2.4m}.  It has
$M_{NS}=1.4\,M_{\odot}$, $M_c=0.167\,M_{\odot}$, $i=75^{\circ}$, and 
$T_{\rm eff}=4230$~K.  Figure~\ref{fig:psrj1628_1.3m} shows the
same model parameters fit to the previous year's data from the 1.3m,
except that $T_{\rm eff}$ is reduced by 100~K to 4130~K in 
order to match the overall lower flux.
The difficulties in fitting presented by the asymmetries described in
Section~\ref{f:obs1} are most evident in Figure~\ref{fig:psrj1628_2.4m},
but also to some extent in  Figure~\ref{fig:psrj1628_1.3m}.
Since the $\approx0.1$ mag difference between the two minima is
accounted for by ellipsoidal modulation, it is not possible to add
heating from a point source without making the model exceed the
observed flux in the phase interval $0.75-1.0$, and especially the peak
at $\phi=1.0$.  Neither ellipsoidal modulation nor axisymmetric pulsar
heating can fit the unequal peaks at $\phi=0.5$ and 1.

We do not enough data to model the \psrb\ system in a similar manner.
In particular, with only one filter, and with no detection of the companion
near inferior conjunction, we don't have a handle on its base temperature
and Roche-lobe filling factor.

\section{Discussion}
\label{f:discuss}

Among the optical light curves of redback MSPs, some are dominated by
pulsar heating and some are not.  In the former group are PSRs
J1023+0038 \citep{thorstensen05}, J2215+5135 \citep{breton13,schroeder14},
J2339$-$0533 \citep{rom11,kong12}, and XSS~J12270$-$4859 \citep{bas14}.
Ones without a dominant heating effect are PSRs J1723$-$2837 \citep{crawford13},
J1816+4510 \citep{kap12,kap13}, and J2129$-$0428 \citep{bellm13}.
This disparity can be understood in terms of the balance of a number
of competing factors, including the intrinsic luminosity of the companion,
the mass ratio and orbital separation, and the spin-down luminosity
of the pulsar.  We have seen some evidence that \psra\ is an intermediate
case, showing mainly ellipsoidal modulation, but possibly some heating.

Here we make a back-of-the-envelope calculation to evaluate whether we should
expect pulsar heating to make an observable contribution to the optical
luminosity of \psra.  The representative model from Section~\ref{f:model}
has a luminosity $L_{\rm bol}=4\pi\,r_L^2\,\sigma\,T_{\rm eff}^4=1.5\times10^{32}$
erg~s$^{-1}$, where $r_L=2.6\times10^{10}$~cm, the radius of the Roche lobe
according to \citet{egg83}.  This value of luminosity is also
obtained by direct summation over the surface of the model star.
The contribution of pulsar irradiation $L_{\rm irr}$ to the optical luminosity
can be expressed in terms of the heating efficiency $\eta$ of an
assumed isotropic pulsar wind as
$L_{\rm irr}=\eta\,\dot E\,r_L^2/4 a^2$, where
$a$ is the orbital separation, and
$\dot E\leq1.3\times10^{34}$ erg~s$^{-1}$.  This is an upper limit
because of the unknown contribution of the kinematic period derivative,
$\dot P_k=P\,v_{\perp}^2/d\,c$,  where $P = 3.2$~ms.  Assuming the
tangential velocity $v_{\perp}\sim100$~km~s$^{-1}$ estimated
in Section~\ref{f:obs1}, $\dot P_k=2.9\times10^{-21}$,
which is 27\% of the observed value, reducing $\dot E$ by
the same percentage.
For the representative model $r_L/a=0.218$,
$L_{\rm irr}=1.1\times10^{32}\,\eta$ erg~s$^{-1}$, and pulsar heating is
comparable to the observed luminosity $L_{\rm bol}$ only if the efficiency
of the process is of order unity, which is not likely for a number
of reasons.  First, the heating may be mediated by an intrabinary shock
between the pulsar
wind and the wind from the companion, which could radiate in all directions.
Second, for stars with deep convective envelopes, as should be the case
here \citep{chen13}, the effective efficiency of re-radiation is
$\lesssim0.5$ \citep{ruc69}.  Third, some of the energy absorbed
goes into the ablated stellar wind.  Due to uncertainties in
the efficiencies of these processes, the true luminosity
and beaming of the pulsar wind, and the distance, we cannot
conclude whether heating should be observed in \psra.

\citet{breton13} estimated an efficiency in the range
$0.1 \le \eta \le 0.3$ for the four systems they studied.
For the nominal distance, the phenomena seen in the light
curve of \psra\ that are {\it not\/} explained by ellipsoidal
deformation could be effects of heating with $\eta$ in this range.
The residual asymmetries may be caused indirectly by the material ablated
from the companion.  If a wind flows off the companion in a cometary
tail that lags the orbital motion, then it is possible that the
intrabinary shock is displaced toward the trailing side of the
companion.  A clear example of asymmetric heating can be seen
in the orbital light curve of PSR J1023+0038, in which the maximum occurs
before phase 0.75.  Both in optical and in X-rays \citep{wou04,bog11},
the trailing side of the companion is brighter than the leading side.

Alternatively, a large starspot could produce an asymmetric
temperature distribution.  Since a starspot would also be associated
with magnetic field intrinsic to the companion, it is also possible
that such a magnetic field could channel the pulsar wind and cause
enhanced local heating, as was speculated by \citet{tan14} to explain
an apparent hot patch on the companion of PSR J1544+4937.  For an
intrinsic large-scale magnetic field to channel the pulsar wind, it must
have a pressure at $\sim 1$ stellar radius above the surface of the companion
at least comparable to the pressure of the relativistic pulsar wind,
i.e., $B^2/8\pi\ge\dot E/4\pi\,(a-2r_L)^2\,c$.
For the representative model, this corresponds to $B\ge12$~G,
a not unreasonable value.

A similar calculation of the efficiency of pulsar heating
can be made for the peak optical luminosity of
\psrb\ at orbital phase 0.75.  Assuming $i\approx90^{\circ}$, and a DM
distance of $0.67$~kpc, we correct the observed $R=21.8$ for
$A_R\approx0.1$.  Converting to flux density using
$f^{\rm Vega}_{\lambda, \rm eff}=2.15\times10^{-9}$ erg~cm$^{-2}$~s$^{-1}$~\AA$^{-1}$
\citep{fuk95}, the flux density at $\lambda_{\rm eff}=6410$~\AA\ is
$f_\lambda=4.5\times10^{-18}$ erg~cm$^{-2}$~s$^{-1}$~\AA$^{-1}$.
If half of the star is illuminated by the pulsar wind
and is re-radiating, its luminosity is $L_{\rm bol}=1.1\times10^{30}$
erg~s$^{-1}$ if $T_{\rm eff}=5000$~K, or $L_{\rm bol}=1.6\times10^{30}$
erg~s$^{-1}$ if $T_{\rm eff}=10,000$~K.  For $M_{NS}=2\,M_{\odot}$,
we have $M_c=0.0305\,M_{\odot}$, and $r_L/a=0.157$.  If the companion
fills its Roche lobe, then $L_{\rm irr}=3.2\times10^{32}\,\eta$ erg~s$^{-1}$.
Black widow pulsars do not necessarily fill their Roche lobes, but
there is room for this one to underfill by a large factor and
still shine by reprocessed pulsar irradiation even if the efficiency
for that process is $\sim 1\%$.

\section{Conclusions and Future Work}
\label{f:conclude}

We have discovered the optical companions of two {\it Fermi}
millisecond pulsars and confirmed the identifications based
on their orbital photometric variability.
Our $R$-band light curve of \psrb\ displays an amplitude
of at least 2.4 mag, with a maximum at superior conjunction of the
companion.  \psrb\ appears to be a classic black widow system.
Its companion is a faint, probably sub-stellar object around
an MSP of high $\dot E$, which explains why its light is
dominated by pulsar heating.  The efficiency of this process is
required to be only $\sim 1\%$. 
Deeper optical observations and multicolor
light curves are needed before this system can be modelled, because 
we have not detected the unheated side of the secondary.  So we do
not know its effective temperature or Roche-lobe filling factor.

The optical modulation of \psra\ is very different.
It is characterized by two peaks and two troughs per orbit,
whose phase relationship to the pulsar ephemeris clearly establishes
that ellipsoidal distortion is the cause of the variation.  The
amplitude of modulation is consistent with a near Roche-lobe
filling secondary star in a high-inclination orbit, which we
conservatively estimate as $i>55^{\circ}$.  This limits the
mass of the secondary to $0.16<M_c<0.30\,M_{\odot}$.  It is
less dense than a main-sequence star of the same mass.

It is not yet possible to derive system parameters more precisely
for \psra\ because its light curve also displays distortions that
are not characteristic of ellipsoidal variations.
An order-of-magnitude estimate for possible pulsar heating in this
system justifies why this process does not make a major contribution
to the radiation from the photosphere, because it would require an
efficiency of order 1.  But in addition, it is difficult to
to improve the fit to the light curve with even a small amount
of isotropic heating from the pulsar.  Instead off-center heating
is required, or else an intrinsic, asymmetric surface-temperature
distribution.

We used distance constraints from the pulsar DM to limit the
effective temperature of the comparison star of \psra\ to
$3560<T_{\rm eff}<4670$~K.  It should have the optical spectrum of 
a late K star.  Multicolor photometry and spectra are needed to
further investigate the cause of the distortions in the light curve
and pin down the system parameters, including the mass of the NS.
Evidence for an overall increase in brightness of $\sim 0.2$~mag
between 2013 and 2014 is present in our observations.  This raises
the possibility that we are seeing an increase in the photospheric
temperature or Roche-lobe filling factor that may eventually cause
sufficient Roche-lobe overflow to form a partial accretion disk and
shut off the radio pulsar mechanism, as has been observed recently
in several similar redback systems.

\acknowledgements
This work is based on observations obtained at the MDM Observatory,
operated by Dartmouth College, Columbia University, Ohio State University,
Ohio University, and the University of Michigan.
JRT gratefully acknowledges support from NSF grant AST-1008217.
We thank the referee for several helpful suggestions.


\begin{thebibliography}{}
\expandafter\ifx\csname natexlab\endcsname\relax\def\natexlab#1{#1}\fi

\bibitem[{{Alpar} {et~al.}(1982){Alpar}, {Cheng}, {Ruderman}, \&
  {Shaham}}]{alpar82}
{Alpar}, M.~A., {Cheng}, A.~F., {Ruderman}, M.~A., \& {Shaham}, J. 1982, \nat,
  300, 728

\bibitem[{{Archibald} {et~al.}(2009){Archibald}, {Stairs}, {Ransom}, {Kaspi},
  {Kondratiev}, {Lorimer}, {McLaughlin}, {Boyles}, {Hessels}, {Lynch}, {van
  Leeuwen}, {Roberts}, {Jenet}, {Champion}, {Rosen}, {Barlow}, {Dunlap}, \&
  {Remillard}}]{archibald09}
{Archibald}, A.~M., {Stairs}, I.~H., {Ransom}, S.~M., {et~al.} 2009, Science,
  324, 1411

\bibitem[Bassa {et~al.}(2014)]{bas14}
{Bassa}, C.~G., {Patruno}, A., {Hessels}, J.~W.~T., {et~al.} 2014,
\mnras, 441, 1825

\bibitem[Bellm \etal(2013)]{bellm13}
Bellm, E., Djorgovski, S. G., Drake, A., \etal\ 2013, \baas, 221, 154.10

\bibitem[Benvenuto \etal(2014)]{ben14}
Benvenuto, O. G., De Vito, M. A., \& Horvath, J. E. 2014, ApJL, 786, L7

\bibitem[Bogdanov \etal(2011)]{bog11}
Bogdanov, S., Archibald, A. M., Hessels, J. W. T., \etal\ 2011, \apj, 742, 97

\bibitem[Bogdanov \etal(2014a)]{bog14a}
Bogdanov, S., Esposito, P., Crawford, F., \etal\ 2014a, \apj, 781, 6

\bibitem[Bogdanov \etal(2014b)]{bog14b}
Bogdanov, S., Patruno, A., Archibald, \etal\ 2014b, \apj, 789, 40

\bibitem[Breton \etal(2013)]{breton13}
Breton, R.~P., van Kerkwijk, M.~H., Roberts, M.~S.~E., \etal\ 2013,
\apj, 769, 108

\bibitem[Casagrande et al.(2008)]{cas08}
Casagrande, L., Flynn, C., \& Bessell, M.\ 2008, \mnras, 389, 585 

\bibitem[{{Chen} {et~al.}(2013){Chen}, {Chen}, {Tauris}, \& {Han}}]{chen13}
{Chen}, H.-L., {Chen}, X., {Tauris}, T.~M., \& {Han}, Z. 2013, \apj, 775, 27

\bibitem[{{Crawford} {et~al.}(2013){Crawford}, {Lyne}, {Stairs}, {Kaplan},
  {McLaughlin}, {Freire}, {Burgay}, {Camilo}, {D'Amico}, {Faulkner}, {Kramer},
  {Lorimer}, {Manchester}, {Possenti}, \& {Steeghs}}]{crawford13}
{Crawford}, F., {Lyne}, A.~G., {Stairs}, I.~H., {et~al.} 2013, \apj, 776, 20

\bibitem[Cordes \& Lazio(2002)]{cor02}
Cordes, J. M., \& Lazio, T. J. W 2002, arXiv:astro-ph/0207156

\bibitem[Deller \etal(2012)]{del12}
Deller, A. T., Archibald, A. M., Brisken, W. F., \etal\ 2012, ApJL, 756, L25

\bibitem[Eggleton(1983)]{egg83}
Eggleton, P. P. 1983, \apj, 268, 368

\bibitem[{{Eichler} \& {Levinson}(1988)}]{eichler88}
{Eichler}, D., \& {Levinson}, A. 1988, ApJL, 335, L67

\bibitem[Fukugita \etal(1995)]{fuk95}
Fukuguti, M., Shimasaku, K., \& Ichikawa, T. 1995, \pasp, 107, 945

\bibitem[{{Gentile} {et~al.}(2014){Gentile}, {Roberts}, {McLaughlin}, {Camilo},
  {Hessels}, {Kerr}, {Ransom}, {Ray}, \& {Stairs}}]{gentile14}
{Gentile}, P.~A., {Roberts}, M.~S.~E., {McLaughlin}, M.~A., {et~al.} 2014,
  \apj, 783, 69

\bibitem[Geppert \& Urpin(1994)]{gep94}
Geppert, U., \& Urpin, V. 1994, \mnras, 271, 490

\bibitem[{{Halpern} {et~al.}(2013){Halpern}, {Gaidos}, {Sheffield},
  {Price-Whelan}, \& {Bogdanov}}]{halpern13}
{Halpern}, J.~P., {Gaidos}, E., {Sheffield}, A., {Price-Whelan}, A.~M., \&
  {Bogdanov}, S. 2013, ATel, 5514, 1

\bibitem[He \etal(2013)]{he13}
He, C., Ng, C.-Y., \& Kaspi, V. M. 2013, \apj, 768, 64

\bibitem[{{Hessels} {et~al.}(2011){Hessels}, {Roberts}, {McLaughlin}, {Ray},
  {Bangale}, {Ransom}, {Kerr}, {Camilo}, \& {Decesar}}]{hessels11}
{Hessels}, J.~W.~T., {Roberts}, M.~S.~E., {McLaughlin}, M.~A., {et~al.} 2011,
  in AIP Conf. Proc. 1357, Radio Pulsars: An Astrophysical Key to Unlock
the Secrets of the Universe, ed. M.~{Burgay} \etal\ (Melville, NY: AIP), 40

\bibitem[{{Kaplan} {et~al.}(2013){Kaplan}, {Bhalerao}, {van Kerkwijk},
  {Koester}, {Kulkarni}, \& {Stovall}}]{kap13}
{Kaplan}, D.~L., {Bhalerao}, V.~B., {van Kerkwijk}, M.~H., {et~al.} 2013, \apj,
  765, 158

\bibitem[{{Kaplan} {et~al.}(2012){Kaplan}, {Stovall}, {Ransom}, {Roberts},
  {Kotulla}, {Archibald}, {Biwer}, {Boyles}, {Dartez}, {Day}, {Ford}, {Garcia},
  {Hessels}, {Jenet}, {Karako}, {Kaspi}, {Kondratiev}, {Lorimer}, {Lynch},
  {McLaughlin}, {Rohr}, {Siemens}, {Stairs}, \& {van Leeuwen}}]{kap12}
{Kaplan}, D.~L., {Stovall}, K., {Ransom}, S.~M., {et~al.} 2012, \apj, 753, 174

\bibitem[{{Kluzniak} {et~al.}(1988){Kluzniak}, {Ruderman}, {Shaham}, \&
  {Tavani}}]{kluzniak88}
{Kluzniak}, W., {Ruderman}, M., {Shaham}, J., \& {Tavani}, M. 1988, \nat, 334,
  225

\bibitem[Kong \etal(2012)]{kong12}
Kong, A. K. H., Huang, R. H. H., Cheng, K. S., \etal\ 2012, ApJL, 747, L3

\bibitem[{{Landolt}(1992)}]{landolt92}
{Landolt}, A.~U. 1992, \aj, 104, 340

\bibitem[{{Lattimer} \& {Prakash}(2007)}]{lattimer07}
{Lattimer}, J.~M., \& {Prakash}, M. 2007, \physrep, 442, 109

\bibitem[Linares(2014)]{linares14}
Linares, M. 2014, \apj, in press, arXiv:1406.2384

\bibitem[Monet \etal(2003)]{mon03}
Monet, D. G., Levine, S. E., Canzian, B., \etal\ 2003, AJ, 125, 984

\bibitem[{{Papitto} {et~al.}(2013){Papitto}, {Ferrigno}, {Bozzo}, {Rea},
  {Pavan}, {Burderi}, {Burgay}, {Campana}, {di Salvo}, {Falanga},
  {Filipovi{\'c}}, {Freire}, {Hessels}, {Possenti}, {Ransom}, {Riggio},
  {Romano}, {Sarkissian}, {Stairs}, {Stella}, {Torres}, {Wieringa}, \&
  {Wong}}]{papitto13}
{Papitto}, A., {Ferrigno}, C., {Bozzo}, E., {et~al.} 2013, \nat, 501, 517

\bibitem[{{Phinney} {et~al.}(1988){Phinney}, {Evans}, {Blandford}, \&
  {Kulkarni}}]{phinney88}
{Phinney}, E.~S., {Evans}, C.~R., {Blandford}, R.~D., \& {Kulkarni}, S.~R.
  1988, \nat, 333, 832

\bibitem[Pickles(1998)]{pickles98} Pickles, A.~J.\ 1998, \pasp, 110, 863 

\bibitem[Podsiadlowski \etal(2002)]{pod02}
Podsiadlowski, Ph., Rappaport, S., \& Pfahl, E. D. 2002, \apj, 565,1107

\bibitem[Predehl \& Schmitt(1995)]{pre95}
Predehl, P., \& Schmitt, J. H. M. M. 1995, \aap, 293, 889

\bibitem[{{Radhakrishnan} \& {Srinivasan}(1982)}]{radhakrishnan82}
{Radhakrishnan}, V., \& {Srinivasan}, G. 1982, Current Science, 51, 1096

\bibitem[{{Ransom} {et~al.}(2011){Ransom}, {Ray}, {Camilo}, {Roberts}, {{\c
  C}elik}, {Wolff}, {Cheung}, {Kerr}, {Pennucci}, {DeCesar}, {Cognard}, {Lyne},
  {Stappers}, {Freire}, {Grove}, {Abdo}, {Desvignes}, {Donato}, {Ferrara},
  {Gehrels}, {Guillemot}, {Gwon}, {Harding}, {Johnston}, {Keith}, {Kramer},
  {Michelson}, {Parent}, {Saz Parkinson}, {Romani}, {Smith}, {Theureau},
  {Thompson}, {Weltevrede}, {Wood}, \& {Ziegler}}]{ransom11}
{Ransom}, S.~M., {Ray}, P.~S., {Camilo}, F., {et~al.} 2011, ApJL, 727, L16

\bibitem[{{Ray} {et~al.}(2012){Ray}, {Abdo}, {Parent}, {Bhattacharya},
  {Bhattacharyya}, {Camilo}, {Cognard}, {Theureau}, {Ferrara}, {Harding},
  {Thompson}, {Freire}, {Guillemot}, {Gupta}, {Roy}, {Hessels}, {Johnston},
  {Keith}, {Shannon}, {Kerr}, {Michelson}, {Romani}, {Kramer}, {McLaughlin},
  {Ransom}, {Roberts}, {Saz Parkinson}, {Ziegler}, {Smith}, {Stappers},
  {Weltevrede}, \& {Wood}}]{ray12}
{Ray}, P.~S., {Abdo}, A.~A., {Parent}, D., {et~al.} 2012,
  arXiv:1205.3089

\bibitem[Ray \etal(2014)]{ray14}
Ray, P. S., Belfiore, A. M., Saz Parkinson, P., \etal\ 2014, \baas, 223, 140.07

\bibitem[Roberts(2011)]{rob11}
Roberts, M.~S.~E. 2011, in AIP Conf. Proc. 1357, Radio Pulsars:
An Astrophysical Key to Unlock the Secrets of the Universe,
ed. M.~{Burgay} \etal\ (Melville, NY: AIP), 127

\bibitem[{{Roberts}(2013)}]{roberts13}
{Roberts}, M.~S.~E. 2013, in IAU Symp., 291, Neutron Stars and Pulsars:
Challenges and Opportunities After 80 Years,
ed. J. van Leeuwen (Cambridge: Cambridge Univ. Press), 127

\bibitem[Romani(1990)]{rom90}
Romani, R. W. 1990, \nat, 347, 741

\bibitem[{{Romani} {et~al.}(2012){Romani}, {Filippenko}, {Silverman}, {Cenko},
  {Greiner}, {Rau}, {Elliott}, \& {Pletsch}}]{romani12}
{Romani}, R.~W., {Filippenko}, A.~V., {Silverman}, J.~M., {et~al.} 2012, ApJL,
  760, L36

\bibitem[Romani \& Shaw(2011)]{rom11}
Romani, R. W., \& Shaw, M. S. 2011, ApJL, 743, L26

\bibitem[{{Roy} {et~al.}(2014){Roy}, {Bhattacharyya}, \& {Ray}}]{roy14}
{Roy}, J., {Bhattacharyya}, B., \& {Ray}, P.~S. 2014, ATel, 5890, 1

\bibitem[Ruci\'nski(1969)]{ruc69} 
Ruci\'nski, S. M. 1969, AcA, 19, 245

\bibitem[Schlafly \& Finkbeiner(2011)]{sch11}
Schlafly, E. F. \& Finkbeiner, D. F. 2011, \apjs, 737, 103

\bibitem[{{Schroeder} \& {Halpern}(2014)}]{schroeder14}
{Schroeder}, J., \& {Halpern}, J. 2014, \apj, in press, arXiv:1401.7966

\bibitem[{{Stappers} {et~al.}(2013){Stappers}, {Archibald}, {Hessels}, {Bassa},
  {Bogdanov}, {Janssen}, {Kaspi}, {Lyne}, {Patruno}, {Tendulkar}, {Hill}, \&
  {Glanzman}}]{stappers13}
{Stappers}, B.~W., {Archibald}, A.~M., {Hessels}, J.~W.~T., {et~al.} 2013,
  \apj, 790, 39

\bibitem[Tang \etal(2014)]{tan14}
Tang, S., Kaplan, D. L., Phinney, S., \etal\ 2014, ApJL, 791, L5

\bibitem[{{Thorstensen} \& {Armstrong}(2005)}]{thorstensen05}
{Thorstensen}, J.~R., \& {Armstrong}, E. 2005, \aj, 130, 759

\bibitem[{{Wang} {et~al.}(2009){Wang}, {Archibald}, {Thorstensen}, {Kaspi},
  {Lorimer}, {Stairs}, \& {Ransom}}]{wang09}
{Wang}, Z., {Archibald}, A.~M., {Thorstensen}, J.~R., {et~al.} 2009, \apj, 703,
  2017

\bibitem[Woudt \etal(2004)]{wou04}
Woudt, P. A., Warner, B., \& Pretorius, M. L. 2004, \mnras, 351, 1015

\end{thebibliography}

\end{document}